\documentclass[10pt, conference]{IEEEtran}

\usepackage{cite}
\usepackage{amsmath,amssymb,amsfonts}
\usepackage{graphicx}
\usepackage{textcomp}
\usepackage{mathtools}
\usepackage{pgfplots}
\usepackage[hidelinks]{hyperref}
\usepgfplotslibrary{groupplots}
\usetikzlibrary{matrix, fit}
\pgfplotsset{compat=1.17}

\begin{document}

\title{Making Memristive Processing-in-Memory Reliable}

\author{
\IEEEauthorblockN{Orian Leitersdorf, Ronny Ronen, and Shahar Kvatinsky} \IEEEauthorblockA{\emph{Viterbi Faculty of Electrical and Computer Engineering, Technion -- Israel Institute of Technology, Haifa, Israel}} \IEEEauthorblockA{orianl@campus.technion.ac.il, ronny.ronen@technion.ac.il, shahar@ee.technion.ac.il}}


\IEEEoverridecommandlockouts
\IEEEpubid{\begin{minipage}{\textwidth}\vspace{40pt}\ \\[12pt] \centering \copyright 2021 IEEE. Personal use of this material is permitted.  Permission from IEEE must be obtained for all other uses, in any current or future media, including reprinting/republishing this material for advertising or promotional purposes, creating new collective works, for resale or redistribution to servers or lists, or reuse of any copyrighted component of this work in other works.
\end{minipage}} 

\maketitle

\begin{abstract}
Processing-in-memory (PIM) solutions vastly accelerate systems by reducing data transfer between computation and memory. Memristors possess a unique property that enables storage and logic within the same device, which is exploited in the memristive Memory Processing Unit (mMPU). The mMPU expands fundamental stateful logic techniques, such as IMPLY, MAGIC and FELIX, to high-throughput parallel logic and arithmetic operations within the memory. Unfortunately, memristive processing-in-memory is highly vulnerable to soft errors and this massive parallelism is not compatible with traditional reliability techniques, such as error-correcting-code (ECC). In this paper, we discuss reliability techniques that efficiently support the mMPU by utilizing the same principles as the mMPU computation. We detail ECC techniques that are based on the unique properties of the mMPU to efficiently utilize the massive parallelism. Furthermore, we present novel solutions for efficiently implementing triple modular redundancy (TMR). The short-term and long-term reliability of large-scale applications, such as neural-network acceleration, are evaluated. The analysis clearly demonstrates the importance of high-throughput reliability mechanisms for memristive processing-in-memory.
\end{abstract}


\section{Introduction}
\label{section:introduction}

\IEEEpubidadjcol

Emerging processing-in-memory (PIM) technologies possess ample potential for massive computational parallelism. By nearly eliminating the CPU-memory data transfer (\emph{memory wall})~\cite{DarkMemory}, PIM is capable of vastly accelerating computing systems~\cite{NDP}. The traditional memory read/write interface is supplemented with in-memory operations that perform logic on stored data without explicit read/write. 

The memristor~\cite{Memristor} enables true processing-in-memory as a fundamental device that supports both storage and logic. The resistance of a memristor may represent binary information (low resistance for logical 1, high resistance for logical 0). Crucially, the unique property of the memristor enables modifying resistance with an applied voltage. For memristors stored in a crossbar array structure, this unique property enables computing \emph{stateful logic} between the memristors in the crossbar without explicitly reading/writing the memristor values~\cite{mMPU}. As the same memristors are responsible for both storage and logic, memristive processing-in-memory is inherently supported. 

Stateful logic within crossbar arrays supports massive inherent parallelism, which is utilized in the memristive Memory Processing Unit (mMPU) for efficient \emph{high-throughput} operations. Stateful logic techniques include IMPLY~\cite{borghetti2010memristive}, MAGIC~\cite{MAGIC}, and FELIX~\cite{FELIX}. These techniques support gates such as NOT/NOR, NAND, OR and Minority3 within rows/columns of crossbar arrays. They also inherently support parallelism: the same in-row (in-column) gate can be repeated across all rows (columns) with the exact same latency. Therefore, the inherent logic capabilities of memristive memory can be utilized for massive parallelism with minimal peripheral overhead. The memristive Memory Processing Unit (mMPU) constructs an entire memory from crossbar arrays that each support this parallelism simultaneously~\cite{mMPU}, achieving high-throughput operation. Through a sequence of basic PIM gates, the unit is capable of advanced logic operation, such as matrix-vector multiplication~\cite{FloatPIM, MultPIM, abstractPIM} and image convolution~\cite{IMAGING, FloatPIM}. 

\IEEEpubidadjcol

Reliability is an open challenge for memristive processing-in-memory. Soft-errors are categorized into those that alter memristor states over time (indirect), and those that lead to incorrect operations (direct). The former are traditionally addressed via the error correcting code (ECC) technique, while the latter are addressed through techniques such as triple modular redundancy (TMR)~\cite{Shooman}. Processing-in-memory is especially vulnerable to these errors as they may propagate undetected through future computations, without the data being explicitly read. Previous works that address soft-errors for PIM are throughput-limited as they require periphery that is not compatible with PIM parallelism~\cite{InSituAging, NModularStateful, SIMPLY}. Therefore, we discuss novel techniques for high-throughput solutions that support reliable PIM by exploiting parallelism themselves.

This paper details architectural innovations for mMPU reliability, contributing the following:
\begin{itemize}
    \item \emph{Memristor Soft Errors:} Memristive soft errors are analyzed and categorized in the context of high-throughput memristive processing-in-memory.
    \item \emph{Error-Correction-Codes (ECC):} We detail non-traditional techniques that efficiently support high-throughput mMPU operations by exploiting mMPU parallelism themselves, expanding our previous work~\cite{ParityPIM}.
    \item \emph{Triple-Modular-Redundancy (TMR):} We propose efficient high-throughput in-memory TMR. A trade-off between latency, area  and throughput is presented.
    \item \emph{Case Study -- Neural Network Accelerator:} We evaluate the proposed high-throughput ECC and TMR solutions in a case-study. This provides an insight into large-scale reliability for real high-throughput applications.
\end{itemize}

\begin{figure*}[!t]
\centering 
\includegraphics[width=7.0in]{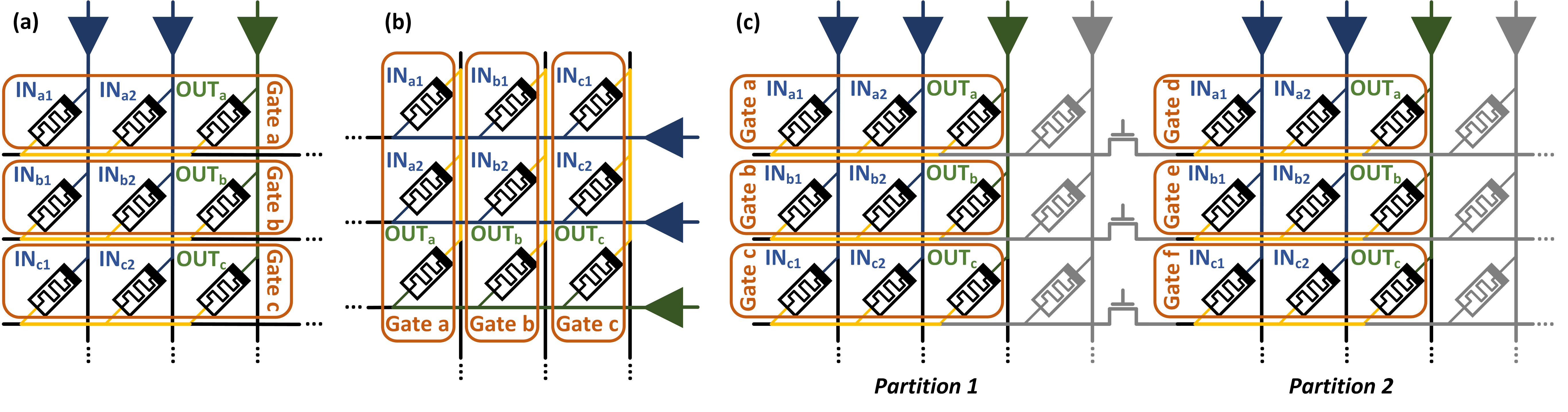}
\caption{MAGIC NOR (a) in-row and (b) in-column operations, each scenario performed simultaneously. The logical inputs are the resistive states of the input memristors prior to the operation, and the final output resistances represent the logical outputs. (c) Transistors may divide the crossbar into partitions that may function as independent computation units, further increasing parallelism by also enabling multiple in-row (in-column) gates within the same row (column).}
\label{fig:crossbar} 
\vspace{-5pt}
\end{figure*}

\section{Background}
\label{sec:background}

\subsection{Stateful Logic}
\label{sec:background:stateful}

Memristive crossbar arrays locate memristors at the crosspoints of vertical bitlines and horizontal wordlines. Each memristor stores binary information through resistance (\textit{i.e.}, high resistance for logical 0, and low resistance for logical 1). Stateful logic considers performing logic operations between the states of the memristors, using the memristors, without explicitly reading/writing their values. This is achieved by exploiting the unique property of the memristor: voltage affects resistance. By applying voltages on the bitlines (wordlines) of the crossbar, it is possible to induce logic between memristors in the same row (column). The resistance of the input memristors prior to the operation represents the logical inputs, and the resistance of the output memristor afterwards represents the logical output. Emerging stateful logic techniques include IMPLY~\cite{borghetti2010memristive}, MAGIC~\cite{MAGIC}, and FELIX~\cite{FELIX}. They support logical gates such as NOT/NOR, NAND, OR and Minority3.

Stateful logic inherently supports row/column parallelism within crossbar arrays. As stateful logic operations are performed by applying voltages along bitlines (wordlines), the same voltages may induce logic in all of the rows (columns) of the crossbar at the same time, as seen in Figures~\ref{fig:crossbar}(a,b). Additionally, partitions may dynamically divide the crossbar array to support multiple in-row (in-column) gates in the same row (column) in parallel~\cite{FELIX}, as shown in Figure~\ref{fig:crossbar}(c).

\subsection{Memristor Soft Errors}
\label{sec:background:softErrors}

Memristors are vulnerable to soft errors, similar to other memory technologies~\cite{SoftErrorTrends}. Soft errors alter the logical state of the memristor (\textit{i.e.}, the resistance) without harming its functionality. Conversely, hard errors permanently damage memristor functionality. Hard errors can be addressed with various testing circuits~\cite{EfficientTesting, ReliableNonVoltatileMemory}, yet soft errors are more difficult to detect. Soft errors in PIM may propagate through consecutive operations without the data ever being explicitly read, while traditional reliability techniques are designed across the read/write interface. Furthermore, the high-throughput mMPU may access/update massive amounts of data at once, which increases the potential for soft errors.
We categorize soft errors into \emph{indirect} and \emph{direct} soft errors.

\subsubsection{{Indirect Soft Errors}}
\label{sec:background:softErrors:indirect}
These soft errors generally occur \emph{over time}, and are divided into the following types:
\begin{itemize}
    \item \emph{Retention and state-drift:} The resistance of a memristor drifts over time, \textit{e.g.}, due to diffusion of oxygen vacancies~\cite{ReliableNonVoltatileMemory, RRAMRefresh, wiefels2021reliability}. Furthermore, read (logic) operations may induce state-drift in the (input) memristors~\cite{SiemonISCAS, VariabilityAwareDesign, MemristorInfluence, UnsafeWriting}.
    \item \emph{Proximity:} When operations on nearby memristors affect this memristor, e.g., read/write disturbance~\cite{ReliableNonVoltatileMemory}.
    \item \emph{Abrupt:} Rare occurrences that change the resistive state at once, e.g., due to ion-strikes~\cite{SoftErrorsMemristorMemory}. Statistically, these typically affect data stored over time.
\end{itemize}

\subsubsection{{Direct Soft Errors}}
\label{sec:background:softErrors:direct}
These soft errors occur at once, due to a failure in a specific operation, such as:
\begin{itemize}
    \item \emph{Write failures:} When a write is unsuccessful~\cite{ECCMemristor}.
    \item \emph{Incorrect logic:} When a stateful logic gate is incorrect, \textit{e.g.}, due to variability in the resistances and properties of the input and output memristors~\cite{MemristorInfluence, InSituAging, UnsafeWriting, UnsafeWriting2, wiefels2021reliability, SiemonISCAS, VariabilityAwareDesign}.
\end{itemize}

\section{Memristive Memory Processing Unit (mMPU)}
\label{sec:mMPU}

The mMPU constructs an entire memory architecture based on memristive stateful logic for high-throughput computation. 

\subsection{Parallelism}
\label{sec:mMPU:parallelism}

The mMPU enables massively high-throughput operation by utilizing three forms of parallelism:
\begin{itemize}
    \item \emph{Row/Column Parallelism:} Logic repetition across multiple rows/columns, as explained in Section~\ref{sec:background:stateful}. Initially, this was utilized to reduce latency~\cite{SIMPLE, UltraEfficient}, yet recent efforts focus on maximizing throughput~\cite{SIMPLER, Ameer, MultPIM, abstractPIM, IMAGING, FloatPIM}. Single-row (single-column) arithmetic is performed within a single row (column), enabling concurrent execution across multiple rows (columns) for vector operations.
    \item \emph{Partition Parallelism:}
    Transistors may dynamically divide the crossbar array into memristive partitions to enable further parallelism~\cite{FELIX}. Partitions within rows (columns) enable concurrent execution of multiple in-row (in-column) logic gates. As a recently emerging technology, partitions appear to possess vast potential~\cite{FELIX, MultPIM, RIME, ParityPIM, alam2021sorting}.
    \item \emph{Crossbar Parallelism:} The mMPU consists of multiple crossbar arrays that can operate in parallel~\cite{Bitlet}.
\end{itemize}

These three parallelism forms enable massive throughput, capable of tremendously accelerating many applications. 

\begin{figure*}[!t]
\centering 
\includegraphics[width=6.3in]{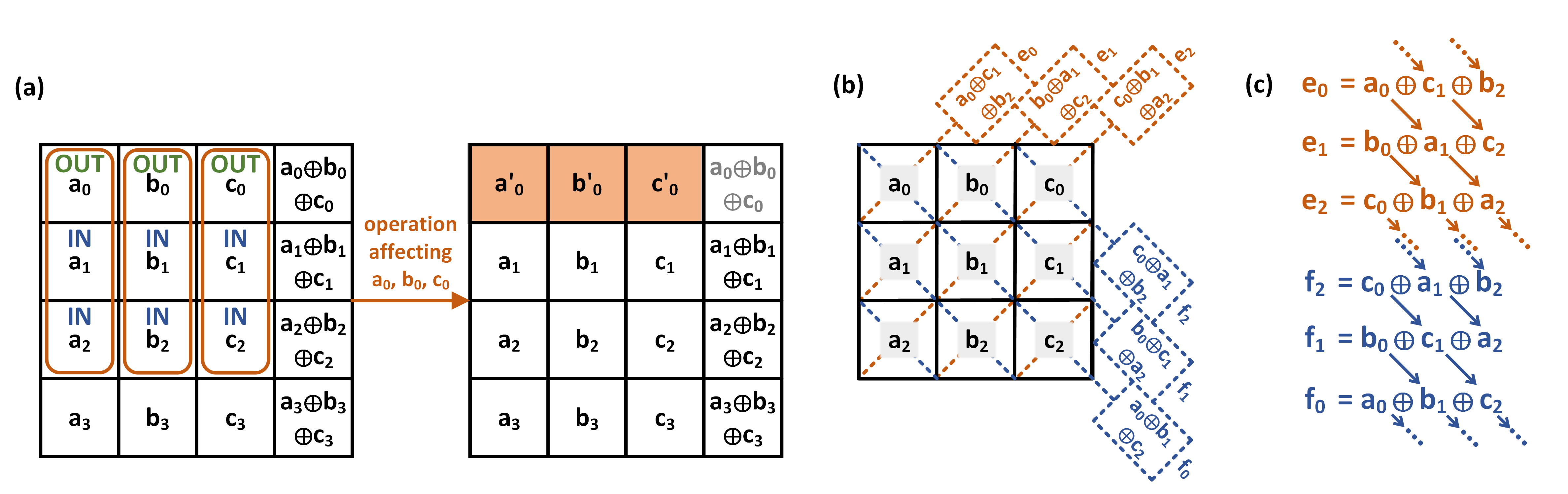} 
\caption{(a) Naive ECC solution that stores parity bits horizontally. An in-column operation is not compatible with this solution as the top-right check-bit has all $n$ of its data-bits updated (in one cycle). Thus, $O(n)$ cycles are required for ECC update. (b) The proposed solution that stores parity along diagonals to enable high-throughput operation in all cases. (c) The shift pattern in the diagonal technique that enables implementation via barrel shifters~\cite{NNPIM, UltraEfficient}.}
\label{fig:proposedSolutions}
\vspace{-10pt}
\end{figure*}

\subsection{Arithmetic Functions}
\label{sec:mMPU:interface}

The mMPU architecture expands the basic stateful logic gates (\textit{e.g.}, NOR) to intra-crossbar arithmetic \emph{functions} (\textit{e.g.}, addition~\cite{FELIX, NishilLogicDesign, IMPLYDesignPrinciples, Bitlet, MultPIM, FloatPIM}, multiplication~\cite{Ameer, RIME, MultPIM}). The mapping of the Boolean functions to in-memory logic gates is performed either manually~\cite{Ameer, RIME, MultPIM} or via automated tools~\cite{SIMPLER, STAR}. The Boolean functions are typically mapped to a single row (column) to allow concurrent execution across multiple rows (columns). This provides vectored operations with low latency, such as vector addition and element-wise vector multiplication. Additional functions, such as matrix-vector multiplication~\cite{FloatPIM, MultPIM, abstractPIM}, are also supported.

The mMPU performs operations at the arithmetic scale, which are utilized for larger applications. The mMPU controller~\cite{mMPUController, CONCEPT} receives instructions from the CPU to perform an arithmetic function (\textit{e.g.}, vector addition) within certain crossbar arrays, and converts the function to stateful logic gates via the aforementioned mapping techniques~\cite{Ameer, RIME, MultPIM, SIMPLER, STAR}. Larger applications utilize these in-memory arithmetic functions with libraries at the CPU-level to enable more complex operations, while still requiring low CPU/memory data transfer as only the instructions are communicated.

\section{Memristive Error-Correcting-Codes}
\label{sec:ECC}

The error correcting code (ECC) technique utilizes redundant data (check-bits) to improve the long-term reliability of the original information (data-bits)~\cite{Shooman}. ECC for the mMPU primarily addresses \emph{indirect} soft errors by protecting the stored data over time. 

The proposed ECC mechanisms are provided on a per-function basis. Each arithmetic function that the mMPU supports utilizes input, intermediate, and output memristors. The proposed solution verifies the correctness of the input data before function execution, and updates the ECC for the output memristors following the execution. This follows from the fact that the function outputs may serve as future inputs. Soft-errors in intermediate memristors are addressed in Section~\ref{sec:TMR}.

Traditionally, ECC is implemented along the memory read/write interface with low throughput~\cite{ECCMemristor}. The lack of such an interface in the mMPU presents the first challenge: data may be accessed and altered within the memory without flowing through the memory interface. Yet, a larger challenge arises from the high-throughput computation supported by the mMPU. As the mMPU is capable of processing vast amounts of data at once (\textit{e.g.} approximately 100 TB/sec for $8192$ crossbars, each sized $1024 \times 1024$, consuming only 1GB of memory~\cite{Bitlet}), the ECC solutions must be capable of \emph{high-throughput} verification and computation. This is achieved by utilizing the mMPU parallelism itself. When an arithmetic function is repeated across all rows (columns), then the inputs/outputs of the function (which require ECC verify/compute) are located across columns (rows) of data. Therefore, high-throughput ECC must support efficient verification/computation for columns/rows of data.

\begin{figure*}[!t]
\centering 
\includegraphics[width=6.5in]{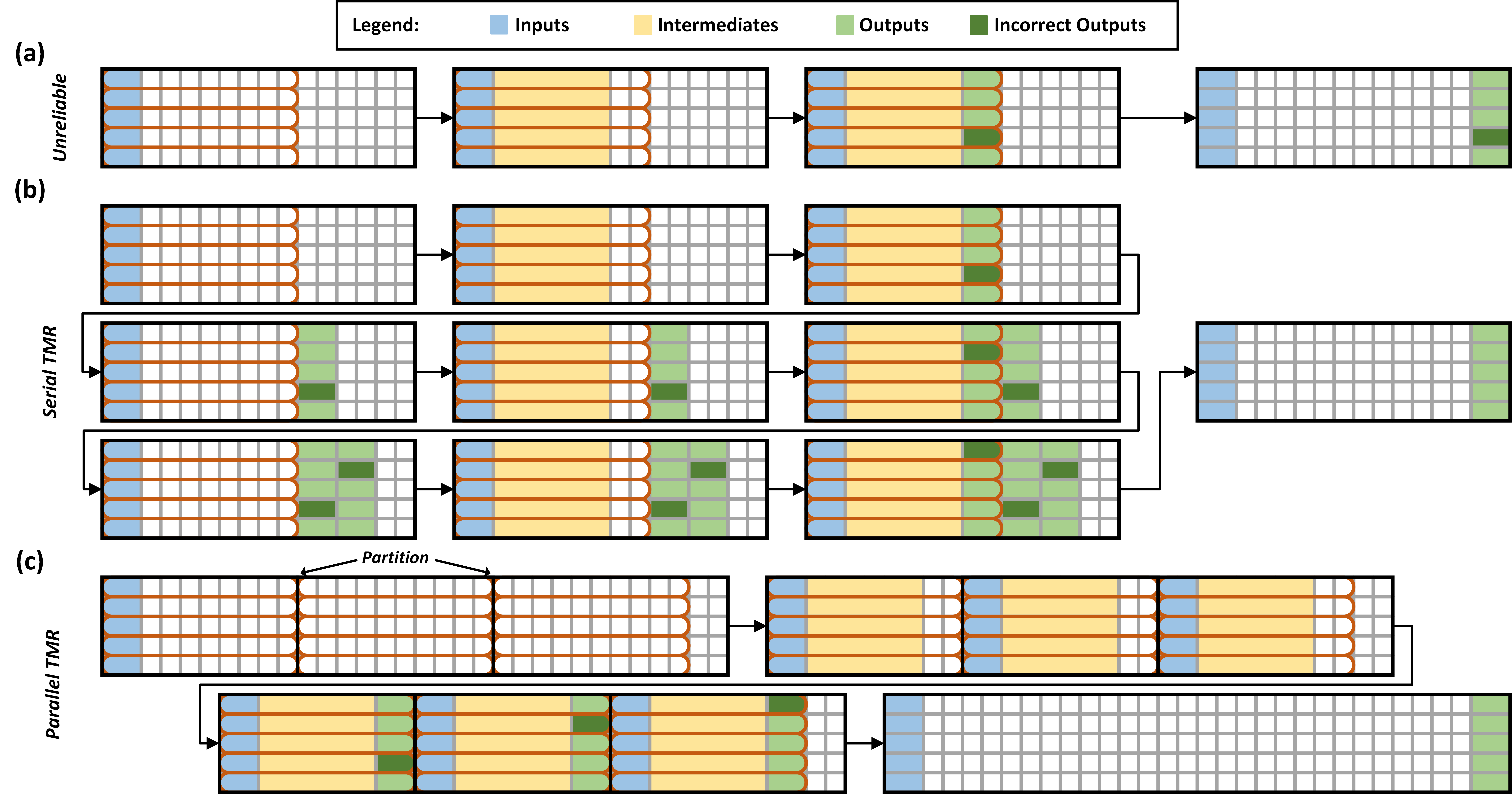} \caption{(a) Unreliable computation of a single-row function in parallel across rows. Reliable computation of the same function through the proposed (b) serial and (c) parallel TMR solutions. }
\label{fig:tmr}
\vspace{-10pt}
\end{figure*}

The naive solution would be to store check-bits within the memory in a horizontal configuration. For example, designating the eighth bit of every horizontal byte as a parity bit, as illustrated in Figure~\ref{fig:proposedSolutions}(a). Consider an in-row logic gate (\textit{e.g.}, NOR) that is repeated along all of the crossbar rows (similar to Figure~\ref{fig:crossbar}(a)). Assuming that the crossbar starts at a valid state (parity bits matching stored data), we seek to reach a valid state that reflects the newly-stored output data. An entire column of information was altered within a single cycle, and the parity bits can be updated in $O(1)$ cycles by utilizing exclusive-or linearity with the same multi-row parallelism (new parity bit can be computed given only old parity bit, old data bit, and new data bit). Unfortunately, when an in-column operation is performed along all columns (similar to Figure~\ref{fig:crossbar}(b)), then $O(n)$ cycles are needed to update the parity-bits, as illustrated in Figure~\ref{fig:proposedSolutions}(a). Therefore, high-throughput ECC is only supported in the naive solution for some of the potential logic and arithmetic operations.

The mMPU-compatible solution stores parity-bits along wrap-around diagonals instead of horizontally/vertically~\cite{ParityPIM}. The diagonal pattern emerges from the unique computation capabilities of the mMPU, enabling $O(1)$ overhead support for all potential user operations. By storing parity-bits along both leading and counter diagonals, single error-correction is achieved, through multi-dimensional parity~\cite{MultidimensionalParity}, per $m\times m$ block in the $n \times n$ crossbar (\textit{e.g.,} $m \approx 16, n \approx 1024$). The diagonal parity-bits are illustrated in Figure~\ref{fig:proposedSolutions}(b) as diagonal extensions to the block (for illustration purposes). The check-bits are stored in a dedicated extension that is also based on memristive memory. The communication between the crossbars utilizes a barrel shifter~\cite{NNPIM, UltraEfficient} to emulate diagonal wires, following the pattern shown in Figure~\ref{fig:proposedSolutions}(c). While the barrel shifter is peripheral, the communication between the crossbars remains stateful (similar to partitions). The dedicated extension may work in parallel to the main memory, enabling moderate latency overhead of $26\%$ on average~\cite{ParityPIM}.

\section{Memristive Triple Modular Redundancy}
\label{sec:TMR}

The triple modular redundancy (TMR) technique improves logic reliability through a simple concept: compute the same function three times and vote between the results~\cite{Shooman}. TMR for the mMPU primarily addresses \emph{direct} soft-errors.

Consider a single-row arithmetic function (\textit{e.g.}, multiplication~\cite{MultPIM}) that is repeated across all of the crossbar rows (\textit{e.g.}, element-wise vector multiplication~\cite{MultPIM}), as illustrated in Figure~\ref{fig:tmr}(a). Errors in the stateful gates of the function, in any of the rows, may\footnote{Some gate errors are masked and do not result in incorrect function output.} cause incorrect output. That incorrect output may propagate erroneous values when used as a future input.

The proposed serial solution increases reliability through high-throughput TMR. The naive solution is repeated three times, using the same inputs and re-using the intermediate memristors. The outputs are stored in three separate copies. At the end, voting is accomplished using the Minority3 logic gate~\cite{FELIX} with parallelism across all of the rows. As shown in Figure~\ref{fig:tmr}(b), while each iteration encountered a soft-error, the final output is correct. This solution increases latency by approximately $3\times$, while area is only slightly increased (as input/intermediate memristors are shared across iterations). 

The proposed parallel solution also utilizes high-throughput TMR, yet reduces the latency overhead with memristive partitions. Rather than performing the naive solution three times in an iterative fashion, the iterations are executed concurrently with memristive partitions~\cite{FELIX}. In this case, inputs and intermediates cannot be shared without compromising partition independence. Hence, while latency remains constant (voting time is negligible due to high-throughput voting using the Minority3 gate), area increases by $3\times$ (as reuse is not possible). Note that a semi-parallel solution that does not utilize partitions is possible by repeating the function computation across additional rows (thereby reducing \emph{throughput} by $3\times$).

As the proposed voting is performed per-bit, effective reliability is increased. That is, incorrect final output requires that at least two copies have errors at the exact same bits. For example, consider a scenario where voting is performed between outputs $1000, 0100, 0010$. Per-element voting would result in an error as no two copies agree on the final output. Conversely, per-bit voting will choose $0000$. Per-bit voting may only increase reliability over per-element voting, as they differ only when per-element voting is undefined.

The proposed TMR solutions efficiently support the high-throughput mMPU. By exploiting the same PIM parallelism as the mMPU for \emph{high-throughput} TMR, the TMR overhead is relatively low, \textit{i.e.}, $3\times$ latency, $1\times$ area (serial) or $1\times$ latency and $3\times$ area (parallel) compared to an \emph{unreliable baseline}. Conversely, solutions that utilize crossbar periphery may not be compatible with PIM parallelism, thereby requiring up to $1024\times$ latency increase (for $1024$ rows)~\cite{InSituAging, NModularStateful}.

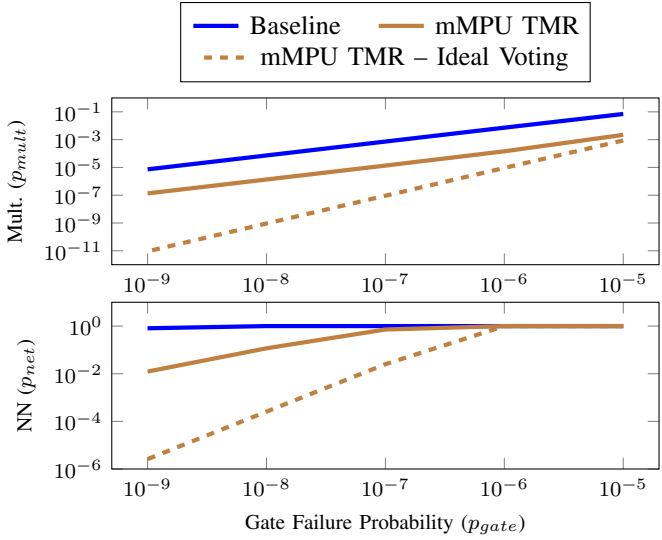
\begin{figure} 
    \centering
  \begin{tikzpicture}
    \begin{groupplot}[
      group style={group size=1 by 2, vertical sep=0.5cm},
      width=\linewidth, height=3.8cm
    ]
    \nextgroupplot[
        xmode=log,
        ymode=log,
        log basis x={10},
        xtick={1e-5, 1e-6, 1e-7, 1e-8, 1e-9},
        ytick={1e-1, 1e-3, 1e-5, 1e-7, 1e-9, 1e-11},
        ymin=1e-12,
        ymax=1e0,
        xmin=1e-9/2,
        xmax=1e-5*2,
        ylabel={\footnotesize Mult. ($p_{mult}$)},
        every axis plot/.append style={ultra thick},
        every tick label/.append style={font=\footnotesize}
    ]
    
    \addplot[color=blue]
    coordinates {(10^(-5), 7.01E-02)(10^(-6), 7.25E-03)(10^(-7), 7.27E-04)(10^(-8), 7.27E-05)(10^(-9), 7.27E-06)}; \label{plot:Baseline}
    
    \addplot[color=brown]
    coordinates {(10^(-5), 2.21E-03)(10^(-6), 1.44E-04)(10^(-7), 1.36E-05)(10^(-8), 1.35E-06)(10^(-9), 1.35E-07)};  \label{plot:TMR}
    
    \addplot[color=brown, dashed]
    coordinates {(10^(-5), 8.75E-04)(10^(-6), 9.18E-06)(10^(-7), 9.20E-08)(10^(-8), 9.20E-10)(10^(-9), 9.20E-12)};\label{plot:IdealTMR}

    

    
    \coordinate (top) at (rel axis cs:0,1);
    
    \nextgroupplot[
        xmode=log,
        ymode=log,
        log basis x={10},
        xtick={1e-5, 1e-6, 1e-7, 1e-8, 1e-9},
        ytick={1e0, 1e-2, 1e-4, 1e-6},
        ymin=1e-6,
        ymax=1e1,
        xmin=1e-9/2,
        xmax=1e-5*2,
        ylabel={\footnotesize NN ($p_{net}$)},
        xlabel={\footnotesize Gate Failure Probability ($p_{gate}$)},
        every axis plot/.append style={ultra thick},
        every tick label/.append style={font=\footnotesize}
    ]

    \addplot[color=blue]
    coordinates {(10^(-5), 1.0000000)(10^(-6), 1.0000000)(10^(-7), 1.0000000)(10^(-8), 0.9999999)(10^(-9), 0.8088945)};
    
    \addplot[color=brown]
    coordinates {(10^(-5), 1.0000000)(10^(-6), 1.0000000)(10^(-7), 0.7180146)(10^(-8), 0.1168216)(10^(-9), 0.0123226)};
    
    \addplot[color=brown, dashed]
    coordinates {(10^(-5), 1.0000000)(10^(-6), 0.9964943)(10^(-7), 0.0252163)(10^(-8), 0.0002554)(10^(-9), 0.0000026)};
    
    
    
    
    \coordinate (bot) at (rel axis cs:1,0);
    
    \end{groupplot}
    
    \path (top|-current bounding box.north)--
          coordinate(legendpos)
          (bot|-current bounding box.north);
    \matrix (table)[
        matrix of nodes,
        anchor=south,
        draw,
        inner sep=0.2em,
        draw,
        minimum height = 13pt
      ] at([yshift=1ex]legendpos)
      {
    \ref{plot:Baseline} Baseline
    \quad \ref{plot:TMR} mMPU TMR\\
    \ref{plot:IdealTMR} mMPU TMR -- Ideal Voting\\};
  \end{tikzpicture}
  \caption{The computation reliability for the unreliable baseline and the proposed TMR, comparing both multiplication failure probability (top) and neural network (NN) failure probability (bottom). At $p_{gate}=10^{-9}$, the baseline network has a soft-error-induced miss-classification rate of $74\%$, while the TMR network has approximately $2\%$ (below the network's inherent accuracy). The proposed TMR with ideal voting is shown in dashed brown.}
  \label{fig:tmrResults} 
  \vspace{-10pt}
\end{figure}

\section{Case Study: Neural-Network Acceleration}
\label{sec:caseStudy}

This section considers the reliability of large-scale applications, assessing the combination of ECC and TMR. We begin by discussing the \emph{computation} reliability of a single fixed-point multiplication, based on the MultPIM algorithm~\cite{MultPIM} and the proposed TMR solutions. We then utilize these results to consider the feed-forward compute reliability of an in-memory neural-network accelerator, such as FloatPIM~\cite{FloatPIM}, under the presence of \emph{direct} soft-errors. Lastly, we analyze the weight degradation over time due to \emph{indirect} soft-errors.

\vspace{-5pt}

\subsection{Multiplication Reliability}
\label{sec:caseStudy:multiplication}

We modify the simulator from MultPIM~\cite{MultPIM} to evaluate the overall reliability provided by the proposed solutions under the presence of \emph{direct} soft-errors. We consider $32$-bit full-precision multiplication according to the state-of-the-art algorithm~\cite{MultPIM}. The simulation accounts for error \emph{masking} in the algorithm. The original simulator involved requests from the algorithm micro-code to perform stateful gates; we inject soft-errors into these requests and measure the logical masking. For the baseline, the MultPIM algorithm is run as usual, and then the final output is compared to the true product. For the proposed solution, the algorithm is performed three times and then voting is performed using in-memory Minority3 (also vulnerable to soft-errors). The probability of multiplication failure, $p_{mult}$, is shown in Figure~\ref{fig:tmrResults} (top) for varying probabilities of single-gate error, $p_{gate}$. Interestingly, the non-ideal voting becomes the bottleneck near $p_{gate}=10^{-9}$, as shown by the dashed line that indicates a \emph{theoretical} ideal-voting solution.

\vspace{-5pt}

\subsection{Neural Network Reliability}
\label{sec:caseStudy:neuralNetwork}

The reliability of a large-scale neural network accelerator is considered, based on the FloatPIM~\cite{FloatPIM} accelerator, the AlexNet~\cite{AlexNet} model ($32$-bit fixed-point), and the ImageNet~\cite{Imagenet} classification dataset. Neural network reliability consists of two aspects: reliable feed-forward \emph{computation} with direct soft-errors, and weight \emph{degradation} due to indirect soft-errors. Both affect the accuracy of the network, while the former affects the short-term accuracy and the latter causes accuracy degradation over time.

\subsubsection{Feed-Forward Computation Reliability}

Feed-forward computation reliability is evaluated based on the multiplication compute reliability, and on previous works that explore error propagation in deep neural networks. Multiplication accounts for the vast majority of the computation in the FloatPIM implementation of AlexNet; thus, we focus only on multiplications. The model requires $M=612 \cdot 10^6$ multiplication per sample. 

The network has inherent logical masking characteristics, thus incorrect multiplications may have no effect on the final classification. We consider the soft-error propagation analysis performed by G. Li~\textit{et al.}~\cite{DNNErrorProp} as the multiplication compute errors here are analogous to the soft-errors they explore (they affect the same intermediate states of the network). For AlexNet, they find that only $p_{mask} = 0.03\%$ of soft-errors affect the final network classification. Therefore, the probability that a single multiplication causes an incorrect classification is approximately $p_{mask} \cdot p_{mult}$. Assuming independence across multiplication reliability, the probability of incorrect final classification due to a soft error is $1-(1-p_{mask} \cdot p_{mult})^M$. These results are plotted in Figure~\ref{fig:tmrResults}~(bottom) for varying $p_{gate}$. We find that the feed-forward compute error is approximately $2\%$ for $p_{gate} \leq 10^{-9}$ using the proposed mMPU TMR (non-ideal voting). Considering that the neural network itself has classification error $\approx 27\%$, this compute error is negligible. 

\begin{figure} 
    \centering
  \begin{tikzpicture}
    \begin{groupplot}[
      group style={group size=1 by 1, vertical sep=0.5cm},
      width=\linewidth, height=5.8cm
    ]
    \nextgroupplot[
        xmode=log,
        ymode=log,
        xlabel={\footnotesize Time (batches)},
        ylabel={\footnotesize $\mathbb{E} [\text{Num. Corrupted Weights}]$},
        ytick={1e-4, 1e-2, 1e0, 1e2, 1e4, 1e6, 1e8},
        every axis plot/.append style={ultra thick},
        every tick label/.append style={font=\footnotesize},
        legend pos=south east,
        legend entries={$p_{input}=10^{-7}$, $p_{input}=10^{-8}$, $p_{input}=10^{-9}$,}
    ]
    
    \addlegendimage{dotted, black, ultra thick}
    \addlegendimage{dashed, black, ultra thick}
    \addlegendimage{black, ultra thick}
    
    \addplot[color=blue]
    coordinates {(10^3, 2.00E+03)(10^4, 2.00E+04)(10^5, 1.99E+05)(10^6, 1.96E+06)(10^7, 1.71E+07)(10^8, 5.98E+07)(10^9, 6.24E+07)}; \label{plot:ECCBaseline}
    
    \addplot[color=blue, dashed]
    coordinates {(10^3, 2.00E+04)(10^4, 1.99E+05)(10^5, 1.96E+06)(10^6, 1.71E+07)(10^7, 5.98E+07)(10^8, 6.24E+07)(10^9, 6.24E+07)};
    
    \addplot[color=blue, dotted]
    coordinates {(10^3, 1.99E+05)(10^4, 1.96E+06)(10^5, 1.71E+07)(10^6, 5.98E+07)(10^7, 6.24E+07)(10^8, 6.24E+07)(10^9, 6.24E+07)};
    
    \addplot[color=brown]
    coordinates {(10^3, 5.06E-04)(10^4, 5.06E-03)(10^5, 5.06E-02)(10^6, 5.06E-01)(10^7, 5.06E+00)(10^8, 5.06E+01)(10^9, 5.06E+02)}; \label{plot:ECC}
    
    \addplot[color=brown, dashed]
    coordinates {(10^3, 5.09E-02)(10^4, 5.09E-01)(10^5, 5.09E+00)(10^6, 5.09E+01)(10^7, 5.09E+02)(10^8, 5.09E+03)(10^9, 5.09E+04)};
    
    \addplot[color=brown, dotted]
    coordinates {(10^3, 5.09E+00)(10^4, 5.09E+01)(10^5, 5.09E+02)(10^6, 5.09E+03)(10^7, 5.09E+04)(10^8, 5.07E+05)(10^9, 4.89E+06)};   
    
    \coordinate (top) at (rel axis cs:0,1);
    
    \coordinate (bot) at (rel axis cs:1,0);
    
    \end{groupplot}
    
    \path (top|-current bounding box.north)--
          coordinate(legendpos)
          (bot|-current bounding box.north);
    \matrix[
        matrix of nodes,
        anchor=south,
        draw,
        inner sep=0.2em,
        draw
      ] at([yshift=1ex]legendpos)
      {
    \ref{plot:ECCBaseline}& Baseline&[5pt]
    \ref{plot:ECC}& mMPU ECC&[5pt]\\};
  \end{tikzpicture}
  \caption{Expected weight corruption weights for the baseline and the mMPU ECC, at varying times (number of batches) with varying $p_{input}$.
  }
  \label{fig:eccResults} 
  \vspace{-15pt}
\end{figure}
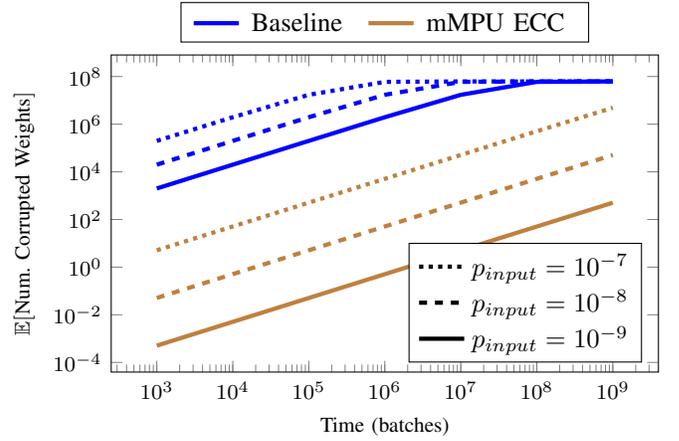

\subsubsection{Weight Degradation}

Weight-degradation is evaluated based on \emph{indirect} soft-errors. As the neural network accelerator constantly accesses all of the $W=62$M weights for every batch, we focus on indirect soft-errors that occur as a result of accessing the memristors. Assuming that accessing a single bit may corrupt that bit with probability $p_{input}$, the probability that a single batch corrupts a 32-bit weight is extrapolated. Then, the probability that that specific weight is corrupted over $T$ consecutive batches is extrapolated, and shown in Figure~\ref{fig:eccResults}. We find that the baseline (no ECC) results in nearly all of the weights corrupted after only $10^7$ batches, while the mMPU ECC maintains an expectation of approximately a single corrupted weight at $10^7$ batches with $p_{input}=10^{-9}$. 

\section{Conclusion and Discussion}
\label{section:conclusion}
The memristive Memory Processing Unit (mMPU) achieves high-throughput operation by exploiting inherent stateful logic parallelism, and by efficiently expanding fundamental logic gates to fast arithmetic functions. Unfortunately, this high-throughput is not compatible with the traditional throughput-limited reliability techniques that address \emph{indirect} and \emph{direct} soft-errors. We detail non-conventional, high throughput, error-correcting-code (ECC) techniques that are designed specifically for the mMPU to address \emph{indirect} errors, as well as high throughput triple-modular-redundancy (TMR) solutions to address \emph{direct} errors. The case study of in-memory neural-network acceleration demonstrates the large-scale implications of the mMPU ECC and TMR, considering both short-term feed-forward accuracy and long-term weight degradation.

The neural network case study presents a unique perspective as the classifier possess massive logical masking capabilities, and as the inherent miss-classification rate is already relatively high. Applications that do not possess these traits, such as database acceleration, will require stronger reliability mechanisms to address memristor soft-errors. Further, such mechanisms may relax the required soft error rates in neural-network acceleration, bridging the gap between the architectural requirements and the physical technology. These mechanisms may include generalizations of TMR, and error-correcting codes with higher correction capability. They must remain high-throughput to efficiently support the mMPU, ideally by utilizing the same mMPU parallelism for reliability.

\vspace{-3pt}

\section*{Acknowledgment}
This work was supported in part by the European Research Council through the European Union's Horizon 2020 Research and Innovation Programe under Grant 757259, and in part by the Israel Science Foundation under Grant 1514/17.

\vspace{-3pt}

\bibliographystyle{IEEEtran}
\bibliography{refs}

\end{document}